\documentclass[fleqn,10pt]{wlscirep}
\usepackage[utf8]{inputenc}
\usepackage[T1]{fontenc}
\usepackage{lineno}
\usepackage{setspace}

\title{Detecting chaos in hurricane intensity}

\author[1,*]{Chanh Kieu}
\author[1,2]{Weiran Cai}
\author[2,3]{Wai-Tong (Louis) Fan}
\affil[1]{Department of Earth and Atmospheric Sciences, Indiana University, Bloomington, IN 47405, USA}
\affil[2]{Department of Mathematics, Indiana University, Bloomington, IN 47405, USA}
\affil[3]{Center of Mathematical Sciences and Applications, Harvard University,  Cambridge, MA 02138, USA}
\affil[*]{Corresponding author: Chanh Kieu (ckieu@indiana.edu)}
\begin{abstract}
Determining the maximum potential limit in the accuracy of hurricane intensity prediction is important for operational practice. Using the phase-space reconstruction method for hurricane intensity time series, here we found that hurricane dynamics contain inherent low-dimensional chaos at the maximum intensity equilibrium. Examination of several chaotic invariants including the largest Lyapunov exponent, the Sugihara-May correlation, and the correlation dimension consistently captures an intrinsic dimension of the hurricane chaotic attractor in the range of 4-5. In addition, the error doubling time is roughly 1-5 hours, which accords with the decay time obtained from the Sugihara-May correlation. The confirmation of hurricane chaotic intensity as found in this study suggests a relatively short limit for intensity predictability of $\sim$18-24 hours after reaching the maximum intensity stage. So long as the traditional metrics for hurricane intensity such as the maximum surface wind or the minimum central pressure is used for intensity forecast, our results support that hurricane intensity forecast errors will not be reduced indefinitely in any modelling systems, even in the absence of all model and observational errors. As such, the future improvement of hurricane intensity forecast should be based on different intensity metric beyond the absolute intensity errors as in the current practice of intensity verification. 
\end{abstract}

\begin{document}
\flushbottom
\maketitle
\thispagestyle{empty}
\doublespacing
%
%
\section*{Main}
Quantifying how far in advance one can predict weather or climate, the so-called atmospheric predictability, is a vital question in real-time forecast. With a wide range of atmospheric systems and operational requirements, there exits however no single method to determine the predictability for all weather phenomena and variables. For example, a large-scale weather system has a typical limit of 2 weeks for geopotential height \cite{Lorenz1969, Lorenz1990, Lorenz1996, Leith1971, metais_and_marcel1986}, yet the predictability for rainfall rate or mesoscale cluster development could be much shorter \cite{Senesi_etal1996, Zhang_etal2003, Durran_etal2013}. Likewise, weather extremes such as tornadoes or convective-scale thunderstorms often cannot be predicted a few hours ahead \cite{HartCOhen2016, Stensrud_etal2009, Bunker_etal2019}. Therefore, a question of what is the maximum time range that one can reliably predict hurricane intensity or track is non-trivial. 

Among many difficulties in understanding hurricane predictability, one central issue roots in the definition of predictability itself. Formally, the predictability of a variable is defined as a maximum time interval beyond which the forecast distribution of that variable becomes indistinguishable from its climatological distribution \cite{Lorenz1969, Shukla, SchneiderGriffies1999, DelSole2004, DelSoleTippett2007}. From this formal definition, it is apparent that predictability must be associated with one specific variable over a given period during which the climatology of the variable is constructed. Thus, predictability is not a universal metric but varies for different variables and different constructions of climatology \cite{DelSoleTippett2007}. 

Given such metric-dependence of predictability, any analysis of hurricane predictability must be therefore carried out for one particular aspect such as track, intensity, decadal shift in the maximum intensity, or seasonal hurricane frequency. A number of recent studies  \cite{KieuMoon2016, kieu_etal2018, Kieu_etal2021} proposed that hurricane dynamics must possess low-dimensional chaos to account for intensity error saturation at 4-5 day lead times as observed in real-time intensity verification. However, a conclusive confirmation for intensity limited predictability still remains open, because real-time forecast errors contain various sources of uncertainties that mask out the intrinsic nature of hurricane intensity predictability.   

Because hurricanes are a complex dynamics system, examining their full dynamics from a strict mathematical perspective is unfeasible at present. This is especially apparent in current numerical models, which contain various nonlinear interactions among different physical parameterizations.  In this study, we use the phase-space reconstruction method in nonlinear dynamics to examine an important question of hurricane intensity predictability limit. By analyzing the output of hurricane intensity from a long hurricane simulation, our ultimate goal is to establish more affirmatively that hurricane dynamics is inherently chaotic at the maximum intensity equilibrium. The ability to state that hurricane intensity has intrinsic chaos is very critical for future hurricane forecast and risk management, as it imposes a strong limit on the intensity predictability that real-time forecast can most achieve. Further quantifying the properties of such intrinsic chaos will eventually allow one to obtain a proper range of hurricane intensity predictability for operational forecasts.
%
%
\section*{Results}
Given the traditional practice of forecasting hurricane intensity based on $V_{MAX}$ and $P_{MIN}$, we apply the phase-space reconstruction method for hurricane intensity in which a time series of these scalars is used to build an intensity phase space. While a single time series of $V_{MAX}$ or $P_{MIN}$ may appear to be too little to explore the complex dynamics of hurricanes, the powerful phase-space reconstruction theorem by Takens \cite{Takens1981} highlights that any single time series should contain rich information about the underlying dynamics if low-dimensional chaos exists. That is, one can explore the main properties of a chaotic attractor for hurricane intensity from any time series, regardless of the output variables \cite{Wolf_etal1985, FraserSwinney1986, Brock1986, Theiler1987, SugiharaMay1990, Casdagli1992, Sugihara_etal1994, WallotMonster2018}. With that, our aim here is to establish that hurricane dynamics contain intrinsic low-dimensional chaos at the maximum intensity limit, which accounts for intensity variability that cannot be eliminated. 

\subsection*{Existence of maximum intensity equilibrium}
Since the phase-space reconstruction method requires a stationary time series \cite{Takens1981, FraserSwinney1986, Brock1986, Theiler1987, SugiharaMay1990, Sugihara_etal1994}, it is necessary to examine first if the maximum intensity equilibrium exists during TC development. In this regard, Figure 1a shows the time series of the maximum surface wind speed obtained from  an 100-day simulation, using the Cloud Model (CM1 \cite{BryanFritsch2002}). One notices in Figure 1a that the model vortex experiences a brief rapid intensification during the first 3-5 days and it quickly settles down to a quasi-stationary state after just 7-9 days into the model integration. These behaviors are typical in hurricane development under idealized conditions as shown in various studies\cite{Hakim2011, Hakim2013, KieuMoon2016}. Note that the existence of a quasi-stationary stage at the maximum intensity equilibrium is also apparent in our simulation, which is consistent with modelling and  theoretical models in recent studies \cite{Kieu2015, KieuWang2017a}. From the practical standpoint, the existence of the stable equilibrium at the maximum intensity state is still an open question due to its sensitivity to model configurations and environmental assumptions \cite{Montgomery_etal2009, Hakim2011,Kieu2015}. However, with the experiment settings described in the Method section, the stable equilibrium of the model maximum intensity (MMI) can be captured and well maintained during the entire 100-day period, thus allowing us to examine the phase-space reconstruction for hurricane intensity as expected.  

Given the MMI equilibrium, it is seen that the maximum intensity does not take one single value but highly fluctuates with time, similar to what obtained in previous studies \cite{Hakim2011, Hakim2013, KieuMoon2016}. 
As shown in Figure 1, temporal fluctuations at the MMI equilibrium is observed not only for the maximum surface wind ($V_{MAX}$) but also for other variables including the minimum central pressure ($P_{MIN}$), the maximum boundary-layer inflow ($U_{MAX}$), and the maximum vertical motion in the eyewall region ($W_{MAX}$). These fluctuations are consistent with  previous studies \cite{Hakim2011, Hakim2013, KieuMoon2016}, which show no apparent difference between chaotic and stochastic variability from the statistical standpoint. This intensity variability highlights an important open question in hurricane dynamics: do these fluctuations reflect the low-dimensional deterministic chaos of hurricane intensity, model random truncation errors, or a manifestation of high-dimensional nonlinearity projection (the so-called process or stochastic noise in \cite{Sugihara_etal1994, Casdagli1992})? 

\begin{figure}[ht]
\centering
\includegraphics[width=14cm]{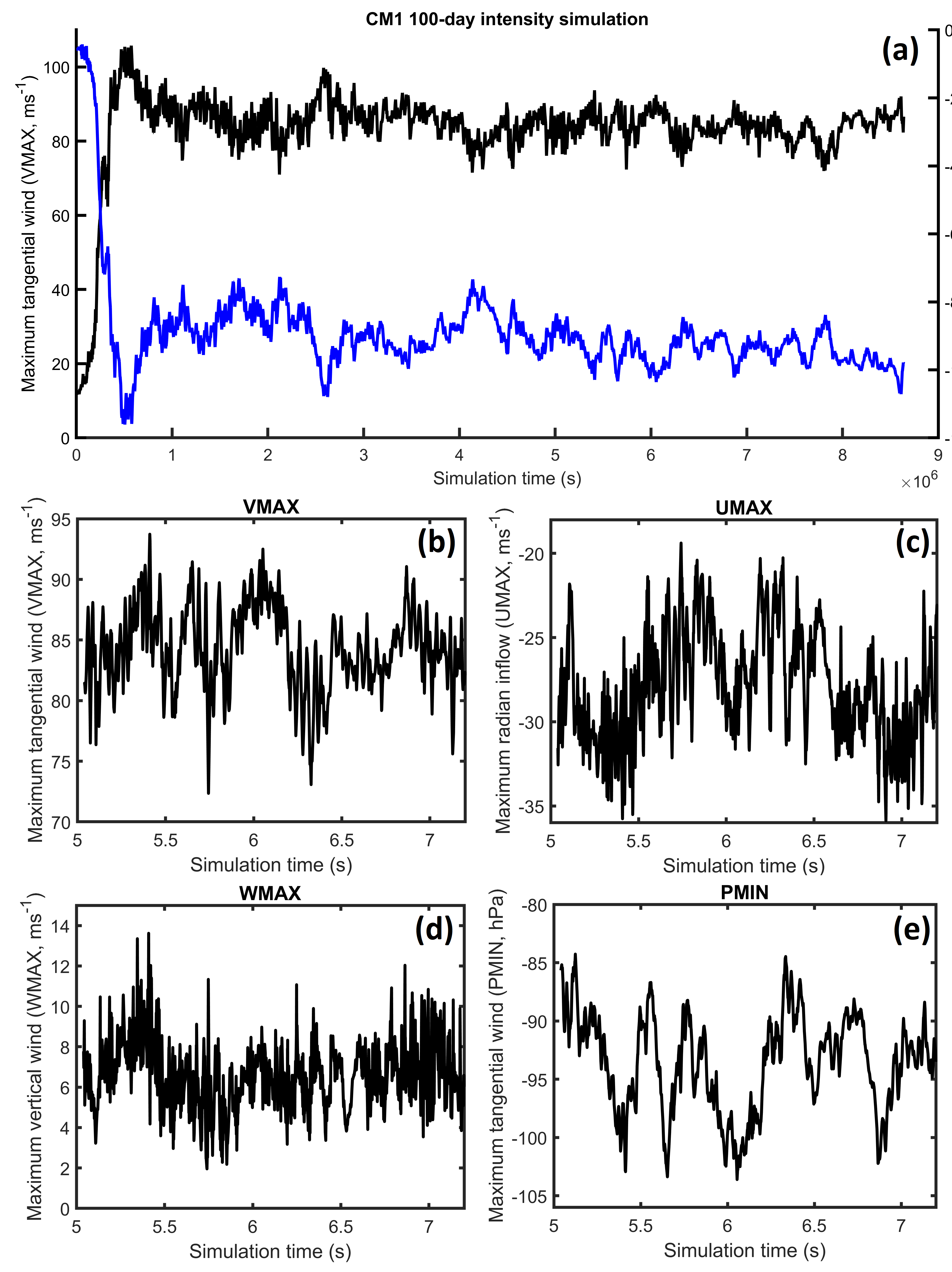}
\caption{(a) Time series of the maximum surface wind $V_{MAX}$ (black, unit $ms^{-1}$) and the minimum surface pressure deficit $P_{MIN}$ (blue, unit hPa) from an 100-day simulations using the CM1 model; (b) A close-up window of the $V_{MAX}$ time series during the maximum intensity equilibrium from day 57-81 of the CM1 simulation; (c)-(d) similar to (b) but for the maximum radial inflow at the surface $U_{MAX}$, the maximum vertical motion in the eyewall region $W_{MAX}$ and $P_{MIN}$, respectively.}
\label{fig1}
\end{figure}

From the time series output, it should be noted that all numerical models appear to be  stochastic \cite{Nguyen_etal2020}. This is because numerical truncation errors can be amplified by nonlinearity and projected onto the time series, resulting in an unexplained noise in the model output \cite{Brock1986, Casdagli1992, Sugihara_etal1994, KantzSchreiber2003}. This stochastic nature of model time series is especially true for modern modelling systems, which employ various stochastic paramterization schemes or random switches such as convective triggering mechanism \cite{Palmer2001, Christensen_etal2015, Dorrestijn_etal2015, Zhang_etal2015, Nguyen_etal2020}. As such, the strong fluctuation of intensity as shown in Figure \ref{fig1} is always present for any model output.

With the existence of the MMI equilibrium as shown in Figure 1, we will examine in the next section three key measures including i) the largest Lyapunov exponent, ii) the Sugihara-May correlation, and iii) the correlation dimension of hurricane intensity. These are the main invariants of chaotic attractors, which can help answer the central question of the potential existence of low-dimensional chaos for hurricane intensity in this study.

\subsection*{Converging largest Lyapunov exponent}
\begin{figure}[ht]
\centering
\includegraphics[width=10cm]{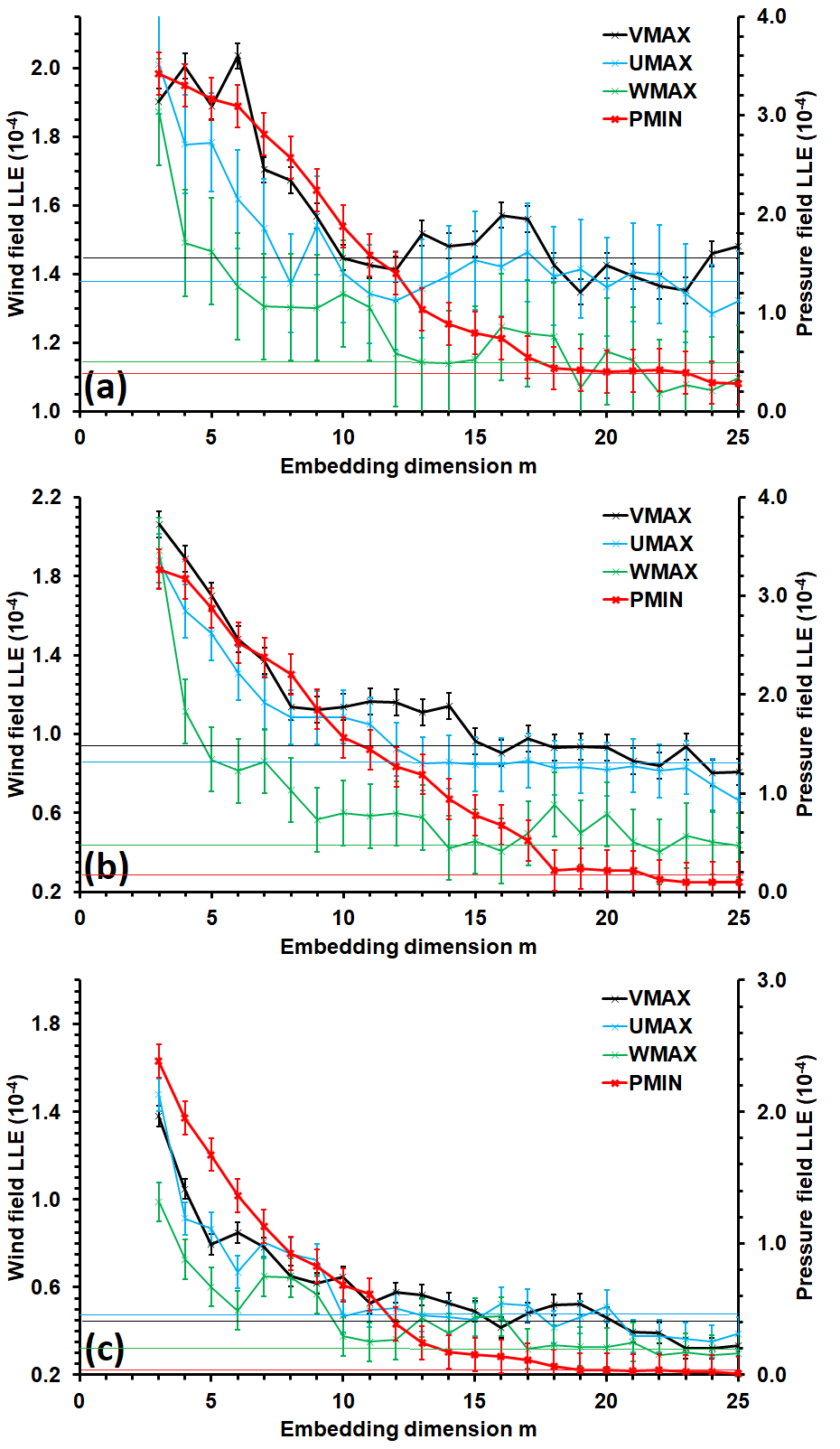}
\caption{Dependence of the largest Lyapunov exponent (LLE, unit $10^{-4} s^{-1}$) on the embedding dimension $m$ for $V_{MAX}$ (black), $U_{MAX}$ (cyan), $W_{MAX}$ (green) and $P_{MIN}$ (red, right axis) for (a) delay time $\tau = 10 \;mins$, (b) $\tau = 30 \;mins$, and (c) $\tau = 45 \;mins$. Error bars denote the 95\% confidence intervals obtained during the maximum intensity equilibrium. Thin solid lines indicate the asymptotic values that LLEs approach when $m$ increases.}
\label{fig2}
\end{figure}

To examine the nature of the variability in the time series $V_{MAX}$, $U_{MAX}$, $W_{MAX}$, and $P_{MIN}$, we show in Figure \ref{fig2} the largest Lyapunov exponent (LLE) $\lambda$ obtained from the these four time series for a range of delay time ($\tau$) between 10-60 minutes. Here, we follow the standard definition of $\tau$as the number of steps between two adjacent data points in a time series. Note that this range of $\tau$ is based on the nature of hurricane dynamics, which is strongly governed by convective activities at a time scale of minutes to hours. Of course, a positive LLE is generally insufficient to conclude whether the variability in a time series is a result of low-dimensional chaos or not. However, the existence of such a positive LLE is a necessary condition that any chaotic system must possess and so we need to examine it first \cite{Wolf_etal1985, FraserSwinney1986, Theiler1987, Brock1986, Sugihara_etal1994}. Details of LLE calculation based on a modified version of the Wolf algorithm by Brock (1986)\cite{Brock1986} are provided in the Method section. 

One notices two important features from the LLE analyses. First, the LLEs derived from all time series display a consistent behavior for all $\tau$ between 10-60 minutes, with a decrease of LLE for a larger embedding dimension ($m$) and a subsequent leveling off in the range of $0.5-1.4\times 10^{-4} s^{-1}$ for $m\ge10$. Recall from the definition of LLE that an LLE of $1\times 10^{-4} s^{-1}$ is equivalent to a doubling time of $\sim$ 3 hr in the full physical dimension. Thus, the range of LLEs shown in Figure \ref{fig2} suggests that an initial error would be doubled every 1-5 hours at the maximum intensity equilibrium. While this is relatively broad range, the fact that all LLEs are positive and convergence towards a stable range when $m$ increases suggests that there exists low-dimensional chaotic for hurricane intensity when the embedding dimension is sufficiently large. Indeed, the decaying of LLEs with $m$ as seen in Figure \ref{fig2} indicates that a too small value for the embedding dimension of the phase space cannot capture a chaotic attractor. As $m$ increases, attractor invariants such as LLEs must converge towards a more stable value, if a low-dimensional chaotic attractor really exists. In this regard, the decay of LLEs with $m$ observed in our time series provides direct clue about the possible existence of intensity chaos.  


Second, Figure \ref{fig2} shows that all LLEs converge towards a stable value for the embedding dimension $m\ge10$, regardless of the time series or time delay values used to reconstruct the phase space. Although the value of the stable LLE cannot be precisely determined but varies between $0.5-1.4\times 10^{-4} s^{-1}$ for different time delays and variables, the fact that such a stable value exists for $m\ge 10$ is important here. Namely, this convergence of LLEs implies that a low-dimensional chaotic attractor of hurricane intensity has an intrinsic dimension $n \approx 4-5$, according to the Takens embedding theorem \footnote{Note that phase space reconstruction generally requires a minimum dimension $m = 2n+1$, where $n$ is the dimension of the attractor such that the invariants of the attractor can be properly estimated. This correlation dimension $n$ is independent of the embedding space dimension $m$ for $m\ge 2n+1$. See a proof in \cite{Brock1986}.}. Of course, finding a proper embedding dimension $m$ from a given time series is often ad-hoc and dependent on other choices of parameters such as time delay, sampling frequency, or sample size. Nevertheless, our sensitivity estimations of $m$ using different methods such as the false nearest neighbor (FNN) method \cite{FraserSwinney1986, Sugihara_etal1994, WallotMonster2018} capture a similar minimum limit for $m \in [10-14]$. Furthermore, this estimation is consistent among different model outputs. Thus, the intensity chaotic attractor appears to require a minimum embedding dimension $m \sim 10$ for the intensity phase-space reconstruction. 

It is of interest to note also from Figure \ref{fig2} that the LLE obtained from CM1 model output seems to be quite different between the wind (i.e., $U_{MAX}$, $V_{MAX}$, and $W_{MAX}$) and the pressure (i.e., $P_{MIN}$) time series. Specifically for the CM1 simulations herein, LLE is $\sim 0.5-1.4 \times 10^{-4} s^{-1}$ for $V_{MAX}$, $U_{MAX}$, or $W_{MAX}$, but it is noticeably smaller ($\sim 0.2-0.5 \times 10^{-4} s^{-1}$) for $P_{MIN}$. In addition, the convergence of LLE with $m$ for the $P_{MIN}$ time series occurs for $m\ge16$ as compared to the range of 10-14 for the wind time series. This difference between LLEs obtained from the wind and the pressure variables is statistically insignificant in our analyses, yet it may reflect different predictability for different state variables in a multi-scale system with the co-existence of fast and slow-varying processes \cite{Shukla, Goswami_etal1997, Lorenz1992, DelSole_etal2017}. Much like the predictability of rainfall is different from that of temperature or 500-hPa geopotential for some large-scale weather systems, it is possible that the hurricane mass and wind fields possess different predictability ranges. This can help explain why recent studies have proposed to use $P_{MIN}$ as a measure for hurricane intensity in operational forecast instead of $V_{MAX}$, because it potentially allows for more reliable intensity forecast in the long run \cite{Klotzbach_etal20}. 

While our search for the minimum embedding dimension based on the convergence of LLEs differs from other approaches such as the box counting or the correlation dimension method \cite{NicolisNicolis1984, Brock1986, Casdagli1992}, we note that all phase-space reconstruction methods are somewhat subjective and similarly ad-hoc due to the wide range of nonlinear dynamical systems and time series characteristics \cite{KantzSchreiber2003}. Thus, there is always some uncertainty in determining a proper minimum dimension for embedding phase space, which explains why $m$ has a range of [10-16] or LLEs $\sim 0.5-1 \times 10^{-4} \; s^{-1}$ as obtained in our study. Regardless of this uncertainty, the emergence of such a low-dimensional attractor for hurricane intensity with a relatively small value of $m$ is still noteworthy, because a too large embedding dimension would imply that our time series analysis is insufficient to capture chaotic dynamics \footnote{As discussed in \cite{Casdagli1992}, a high-dimensional deterministic chaos would be in fact manifested as stochastic variability, even in the absence of all random noise.}. As a result, the LLE analyses could provide some evidence of low-dimensional chaos for hurricane intensity, at least from the standpoint of error growth on an attractor at the quasi-stationary equilibrium. 

\subsection*{Rapid decay of Sugihara-May correlation}
As discussed in Sugihara et al. \cite{Sugihara_etal1994}, detecting chaos based on the existence of a positive LLE in a time series must be cautioned. This is because the fluctuation in any time series could be manifestation of high-dimension nonlinearity or random noise. One could indeed have a non-chaotic system with a positive LLE if there is sufficiently large random noise in the time series \cite{Brock1986, Casdagli1992, Sugihara_etal1994}. As such, a positive LLE as shown in Figure \ref{fig2} is generally insufficient to guarantee the existence of low-dimensional chaos.    

\begin{figure}[ht]
\centering
\includegraphics[width=10cm]{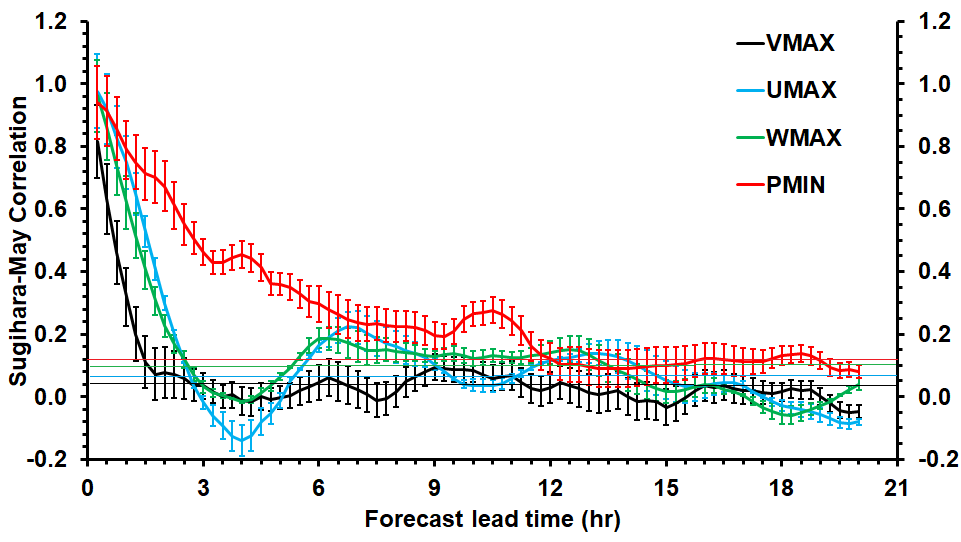}
\caption{Dependence of the Sugihara-May correlation (SMC) on the forecast lead time $T$ for $V_{MAX}$ (black), $U_{MAX}$ (cyan), $W_{MAX}$ (green) and $P_{MIN}$ (red). Error bars denote the 95\% confidence intervals obtained during the maximum intensity equilibrium.}
\label{fig3}
\end{figure}

To further examine the potential existence of low-dimensional chaos in hurricane intensity time series, Figure \ref{fig3} shows the Sugihara-May correlation (SMC) as a function of forecast lead time $T$ for all four variables. Here, SMC is obtained by using a modified version of Sugihara and May's original algorithm, in which the forecast scheme is based on an ensemble average instead of a weighted sum \cite{SugiharaMay1990} or regression method \cite{Casdagli1992} as described in the Method section. Note also that a fixed embedding dimension $m=10$ and the delay time $\tau = 30$ minutes are chosen for this SCM calculation, based on the results from the LLE analyses in the previous section. 

One notices in Figure \ref{fig3} that all SMCs from four different variables show rapid decay with forecast lead time. As discussed in \cite{SugiharaMay1990}, this type of decaying correlation is an inherent characteristic of chaotic dynamics, which is distinct from the random noise variability whose SMC is statistically constant. Of further interest in Figure \ref{fig3} is the consistency of such decaying SMC among all time series, which confirms the limited predictability for hurricane intensity due to low-dimensional chaos, irrespective of model output. Specifically for our CM1 simulation, we observe that SMC decreases from 1.0 to about 0.1 after reaching a limit $T^* \approx$ 2-3 hours for the wind field. Such a chaotic decorrelation time is also consistent with the predictability range obtained from the hurricane-scale dynamics framework in \cite{KieuMoon2016}, which is based on the largest finite Lyapunov exponent in the phase space of hurricane scales. 

Similar to the LLE analyses, it is of significance to mention that the time series for the wind components ($U_{MAX}$, $V_{MAX}$, $W_{MAX}$) display a consistent range of predictability among themselves ($T^* \approx$ 2-3 hours), while $P_{MIN}$ tends to capture a longer decorrelation time ($T'^* \approx$ 12-13 hours) as shown in Figure \ref{fig3}. Such a longer decorrelation time in the $P_{MIN}$ time series again suggests that the pressure field may contain different dynamics, which could allow for more reliable intensity forecast at longer lead times. We note that this difference in SMC between the mass and the wind time series is robust for a range of embedding dimension, delay time, model physical options, stochastic forcings, or initial conditions in all analyses, so long as the phase space is properly reconstructed. The fact that both LLE and SMC analyses provide a consistently different behavior between the wind and pressure fields highlights the possible different predictability for hurricane intensity when using $V_{MAX}$ or $P_{MIN}$.    

Our additional sensitivity analyses with different delay time $\tau$ or embedding dimension $m$ show that the SMC curves display consistent decay and level off only when $m \ge 10$, which is comparable with the embedding dimension obtained from the LLE convergence for the wind field or FNN method. For smaller values of $m$, the SMC curve does not posses a monotonic decay but highly fluctuates with forecast lead time. This result confirms once again that the embedding dimension for hurricane intensity phase space must be sufficiently larger before one can attain consistent characteristics of SMC. 

\subsection*{Estimations of the correlation dimension}
\begin{figure}[ht]
\centering
\includegraphics[width=16cm]{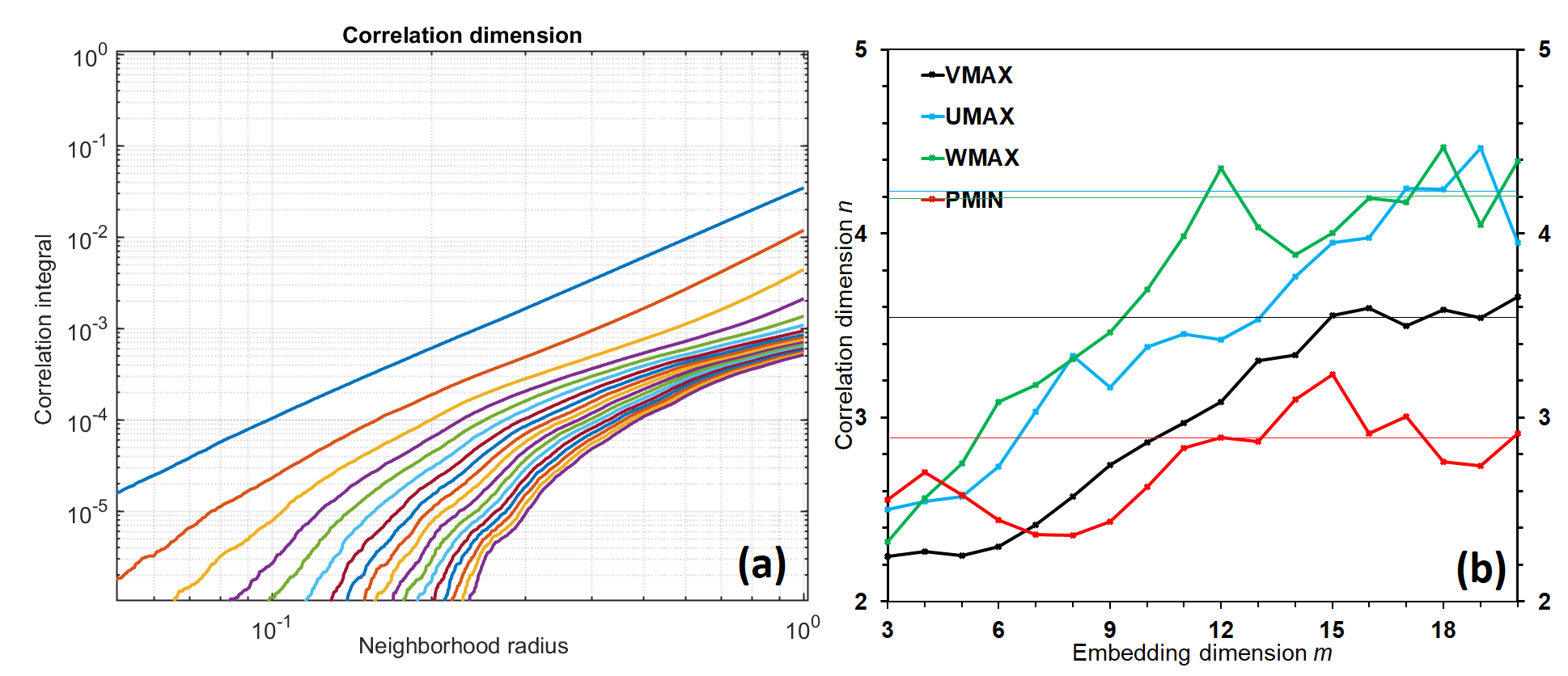}
\caption{(a) Dependence of the correlation integral on the neighborhood radius for different values of embedding dimension $m$, using the Grassberger-Procaccia correlation dimension method for the $V_{MAX}$ time series; and (b) Dependence of the correlation dimension $n$ (i.e., the slope of the correlation integral-radius curve in panel a) on the embedding dimension obtained from $V_{MAX}$ (black), $U_{MAX}$ (cyan), $W_{MAX}$ (green) and $P_{MIN}$ (red) time series.}
\label{fig4}
\end{figure}

The joint convergences of the SMC curve and the LLE for $m \ge 10$ is remarkable, because it indicates the existence of a low-dimensional attractor with a similar intrinsic dimension $n \sim 4-5$. To verify this intrinsic dimension of the intensity chaotic attractor, the Grassberger-Procaccia (GP) correlation dimension algorithm \cite{Theiler1987} is further used to estimate the dimension of the hurricane intensity attractor directly from the CM1 time series (Figure \ref{fig4}a) \footnote{This correlation dimension algorithm is provided as a built-in function in Matlab's Predictive Toolbox.}. While this correlation dimension algorithm has some degree of subjectivity in choosing the best linear fit for correlation integral, these curves do show a saturation slope $n \approx 5-7$, which is the correlation dimension of a chaotic attractor (Figure \ref{fig4}b). Note that this GP correlation dimension is an invariant of a chaotic attractor, which corresponds to a minimum embedding dimension $m \ge 11-15$. Therefore, the slope convergence of the correlation integral curves in Figure \ref{fig4} supports the existence of hurricane intensity chaotic attractor with an intrinsic dimension $n \sim 5-7$, similar to what obtained from the LLE and SMC analyses.

In the search for the intrinsic dimension of hurricane chaotic attractor, we recall a common underlying assumption that possible contributions from random noise must be sufficiently small. This is because noise could strongly interfere with the phase-space reconstruction procedure and result in, e.g, an artificially positive LLE or incorrect correlation dimension estimation \cite{Brock1986, Sugihara_etal1994, Casdagli1992, KantzSchreiber2003}. How random noise impacts our phase-space reconstruction analyses of hurricane intensity is not known.

To address the robustness of our correlation dimension estimation, one could employ different statistical testing methods that could distinguish the difference between chaotic and stochastic time series \cite{Brock1986,BaekBrock1992}. Within the model simulation framework, one can however approach this problem differently by carrying out additional experiments in which random processes in the form of stochastic forcing are included in the CM1 model (see the Method section for details of this stochastic forcing implementation in the CM1 model). Any difference in the estimations of attractor invariants such as LLE, SMC, or correlation between the stochastic and deterministic time series could therefore reveal the role of random noise in the hurricane intensity phase-space reconstruction.     

\begin{figure}[ht]
\centering
\includegraphics[width=10cm]{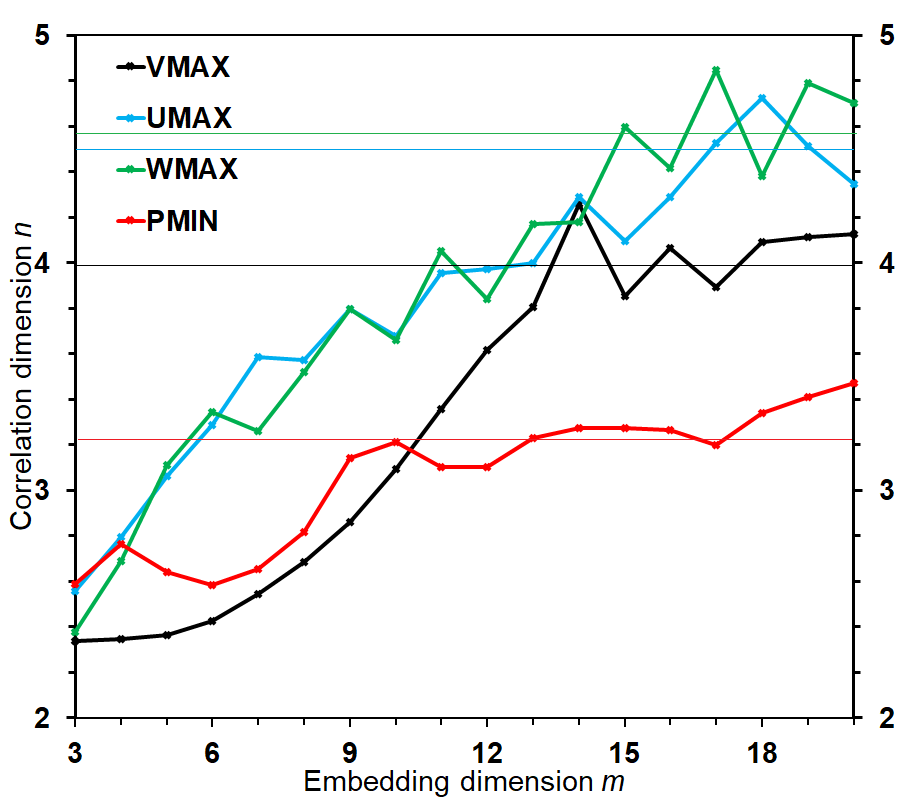}
\caption{Similar to Figure 4b but for the CM1 simulation in which additive random noises  are added to model wind fields at every time step of the model integration over the entire model domain.}
\label{fig5}
\end{figure}

In this regard, Figure \ref{fig5} shows the GP correlation dimension $n$ as a function of the embedding dimension $m$, which is obtained from the CM1 simulation with stochastic forcing implementation. Despite the existence of noise in the intensity time series, one notices apparently from Figure 5 that the correlation dimension displays a consistent behavior among all time series, similar to that obtained from the CM1 deterministic simulations in Figure \ref{fig4}. That is, $n$ increases and levels off for $m>10$. Note however that this consistent behavior is only held for a range of stochastic forcing amplitude. For a sufficiently large value of stochastic forcing, the CM1 model crashes due to violation of the model numerical stability constraint, thus preventing us from examining to what extent random processes would dominate chaotic variability. Similar LLE or SMC analyses for these stochastic simulations confirm the robustness of our phase-space reconstruction (not shown), thus supporting the existence of low-dimensional chaos for hurricane intensity, even in the presence of random noise.         

\section*{Discussion}
From the practical perspective, the values of LLE ($\lambda$), the SMC de-correlation time ($T^*$), and the size of a bounded chaotic attractor ($\Gamma$) are all related, and they together dictate the range of intensity predictability. Indeed, assuming that an initial intensity error is $\epsilon_0$, then the time required to reach the saturation level $\Gamma$, which is often considered as the range of predictability in practical applications, is given by $T_e = \frac{1}{\lambda}ln(\frac{\Gamma}{\epsilon_0})$. If this interpretation of predictability in terms of the saturation time is rational, one would expect that $T_e$ is of the same order of the magnitude as $T^*$. Assume for example $\Gamma \approx 8 m s^{-1}$ from real-time intensity verification \cite{Tallapragada_etal2014a,Tallapragada_etal2015, KieuMoon2016, Bhatia_etal2017,kieu_etal2018}, $\lambda = 1 \times 10^{-4} s^{-1}$, and $\epsilon_0 = 0.5 m s^{-1}$, one obtains $T_e \approx 8 hrs$, which is of the same order of magnitude as $T^*$ obtained from the Sugihara-May's decorrelation time scale (cf. Figure \ref{fig3}). Such consistency thus reveals the nature of chaotic dynamics in determining hurricane intensity predictability as proposed in recent studies \cite{KieuMoon2016,kieu_etal2018}. 

We wish to note that unlike $\lambda$, $T^*$, or $\Gamma$, which can be considered as invariants of a chaotic attractor, the above estimation of $T_e$ depends on the initial condition error $\epsilon_0$. In principle, one could reduce $\epsilon_0$ to as small a value as one wishes (but of course never equal to zero due to the ultimate Principle of Uncertainty) such that $T_e$ can be arbitrarily long \cite{Palmer_etal2014}. However, the logarithm function in the estimation of $T_e$ still imposes a strong constraint on the magnitude of $T_e$ (i.e., a 10-time reduction in $\epsilon_0$ can only lengthen $T_e$ by $\sim$ 2 times). Regardless of how long $T_e$ is, it is eventually the de-correlation time $T^*$ that puts a cut off on the intrinsic predictability of a chaotic system as discussed in Sugihara and May \cite{SugiharaMay1990}, no matter how small $\epsilon_0$ can be reduced. In this regard, the results obtained herein suggests a limited range of 6-12 hours for hurricane intensity predictability, after reaching the maximum intensity stage. 

Regardless of the uncertainties in the range of intensity predictability as obtained from our analyses, the fact that the existence of low-dimensional chaos for hurricane intensity can be confirmed as presented in this study is alone a profound finding. This is because the nonlinear dynamics of hurricanes as well as various physical parameterizations in any hurricane model make it impossible to directly derive any attractor from the governing equations. Therefore, the ability to capture such low-dimensional intensity chaos from the time series of hurricane intensity is important. This is similar to the situation in which one can entirely reconstruct the Lorenz butterfly wing attractor by simply using a single time series output from any state variable, without the need of knowing the full equations of Lorenz's 3-variable model at all. Apparently, the Takens embedding theorem is fundamental here, as it guarantees that the phase-space reconstruction from any time series will be feasible and meaningful if low-dimensional chaos exists. 
%
%
\section*{Concluding remarks}
Determining whether hurricane intensity has limited predictability, and if so, what is the maximum range of hurricane intensity predictability is of importance for operational forecast. In this study, the phase-space reconstruction method was used to probe possible existence of low-dimensional chaos for hurricane intensity. Using the time series output from long hurricane simulations, we presented how potential chaotic behaviors of hurricane dynamics could be obtained from these time series of the model output. 

With the time series of wind and pressure variables extracted at the model maximum intensity equilibrium, here we found that hurricane intensity possesses indeed low-dimensional chaos from several perspectives. Specifically, our analyses of the largest Lyapunov exponent (LLE) and the Sugihara-May correlation (SMC) revealed a consistently positive LLE and a decaying SMC when the embedding dimension of the phase space $m\ge 10$ as expected for chaotic systems. For LLE, all estimations converge towards a divergent rate in the range of $\sim 0.5-1 \times 10^{-4} s^{-1}$, which corresponds to the e-folding time of $\sim$ 3-6 hours. Similarly, the SMC curve shows a consistent decaying of the predicted correlation after $\sim 1-5 \times 10^4 s$, regardless of the presence of random noise. These results together advocate that the variability in hurricane intensity time series is governed by chaotic dynamics, rather than pure stochastic fluctuation or projection of high-dimensional nonlinearity.       

While the LLE and SMC measures depend on a certain choice of embedding dimension thresholds, model resolution, sampling frequency, or phase-space construction methods, we note that our estimations of LLE and SMC are robust for a range of sensitivity analyses. In particular, the convergence of LLE and SMC is consistent among the time series of all wind components and the minimum central pressure. Note, nevertheless, that the estimations of LLE and SMC from the time series of the minimum central pressure provide somewhat a smaller value for LLE and a longer decorrelation time for SMC, as compared to those obtained from the time series of the wind components. This appears to be a notable property of hurricane dynamics,  because it suggests then that the wind and the mass fields tend to have a different range of predictability. The fact that a smaller LLE and a larger SMC time obtained from the mass field, in this regard, indicates that hurricane intensity would have a longer range of predictability if the minimum central pressure is used for intensity forecast. Despite such difference between the mass and wind fields, the predictability range for hurricane intensity appears to be around 6-12 hours once hurricanes attain their quasi-stationary stage. The results provide concrete evidence about hurricane chaotic dynamics, and indicates that any future improvement of intensity accuracy should be based on different intensity metrics beyond the absolute intensity errors, regardless of how perfect our modelling system or observational networks would be in the future.  

As noted in previous studies \cite{KieuMoon2016, kieu_etal2018}, any practical interpretation for hurricane intensity predictability as obtained from LLE/SMC should be cautioned. First, this estimation is only applied to hurricane maximum intensity stage such that the stationary time series can be well maintained under fixed environmental conditions (i.e., the dynamics is already on a chaotic attractor \cite{Brock1986, KantzSchreiber2003, Alligood_etal}). This restriction limits one from examining the variability of hurricane intensity during the early stage of development or as a function of environment. How the intensity predictability of a hurricane depends on its track or different intensity metrics beyond the few scalar variables used in this study is therefore still elusive.      

Second, even if our LLE estimation is accurate, the usefulness of LLE in estimating the predictability range for hurricane intensity is still somewhat limited. This is because one must know both an initial intensity error as well as the saturated error limit such that the range of predictability can be estimated. While the current initial error magnitude is known about 5 knot ($\sim 2.5 m s^{-1}$), our knowledge of the intensity error saturation limit is inadequate at present. The most recent estimation of this intensity error saturation in an axisymmetrical model is $\sim$ 6-8 $m s^{-1}$ \cite{Hakim2011, Hakim2013, KieuMoon2016}, yet similar estimations using full-physics model capture a range of 2-4 $m s^{-1}$\cite{kieu_etal2018, Kieu_etal2021}. Furthermore, this range could change from basin to basin, depending on environmental conditions. As a result, a good estimation for the range of hurricane intensity predictability in the presence of low-dimensional chaos needs to be further examined in the future.    
%
%
\section*{Methods}
\subsection*{Phase-space reconstruction}
In a strict mathematical sense, the governing equations for hurricanes are not closed due to our incomplete understanding of TC dynamics and thermodynamics. As a consequence, all current representations of hurricane processes in numerical models must employ empirical parameterizations that only approximate the true hurricane physics. These physical parameterizations generally contain various uncertainties and simplifications, which prevent one from fully understanding hurricane development. 

Early works by Takens and many others \cite{Takens1981, Brock1986, Theiler1987, SugiharaMay1990, Sugihara_etal1994, Casdagli1992} have shown, however, that the dynamics of a nonlinear system can be reconstructed from a single time series of a state variable under some specific conditions, even in the absence of complete governing equations for the system. Assuming that a nonlinear system possesses low-dimensional chaos at its statistically stationary state, it is possible to examine multidimensional phase portraits of a chaotic attractor by reconstructing the attractor in the phase space of time-lagged coordinates. With this phase-space reconstruction, different invariants of the original chaotic attractor can be effectively obtained if the embedding dimension and time delay are properly chosen.  

There are a range of techniques that have been proposed to find a proper embedding dimension and time delay for phase-space reconstruction such as the averaged mutual information, auto-regression, or false nearest neighborhood \cite{FraserSwinney1986, Sugihara_etal1994, WallotMonster2018}. These methods all share a common principle that basic invariants of a chaotic attractor must be intrinsic, regardless of the reconstruction methods if the embedding dimension and time lag are correct. Among several approaches to detect chaos in a phase space reconstructed from a time series, we present in this study two specific methods based on the estimation of the largest Lyapunov exponent (LLE) and Sugihara and May (1990)'s correlation (SMC) curve. 

For the LLE calculation, an early algorithm for computing LLE from a given time series was proposed by Wolf et al. \cite{Wolf_etal1985}, which has been  improved in many subsequent studies \cite{Rosenstein1997, Kantz1994, Balcerzak_etal2018, Awrejcewicz_etal2018}. For our implementation of the LLE algorithm, a version of Wolf's algorithm presented in Brock \cite{Brock1986} was chosen herein because of its efficiency. The basic steps of Brock's scheme are summarized below (see the full proof of convergence in \cite{Brock1986}):
\begin{itemize}
\item Step 1: construct a set of $m$-history $a^m_t, t=1...N_m\equiv N-(m-1)\tau$ from a given time series $\{a_i\}, i=1...N$ with a given time delay $\tau$ and an embedding dimension $m$; 
\item Step 2: Initialize an error growth cycle by finding the nearest neighborhood $a_{t_1}^m$ of the first m-history $a_1^m$ such that $a_{t_1}^m \neq a_1^m$;
\item Step 3: Choose a prescribed evolution window $q$ and compute $g_1(q)=d_2{(1)}/d_1{(1)}$, where $d_1^{(1)} = \| a_{t_1}^m - a_1^m \|$ and $d_2^{(1)} = \| a_{t_1+q}^m - a_{1+q}^m \|$ are distances in the reconstructed phase space with a given metric $\|\cdot \|$;   
\item Step 4: Perform a loop from $k=2$ to $K=max\{k\,|\,1+kq\le N_m\}$ that repeatedly does the following two main tasks:
\begin{enumerate}
\item Find an index $t_k$ of $t$ to minimize a penalty function $p(a_{t}^m - a_{1+(k-1)q}^m, a_{t_{k-1}+q}^m - a_{1+(k-1)q}^m)$ defined as follows:
\begin{eqnarray}
p(a_{t}^m - a_{1+(k-1)q}^m, a_{t_{k-1}+q}^m - a_{1+(k-1)q}^m) = \| a_{t}^m - a_{1+(k-1)q}^m \| + w|\theta(a_{t}^m - a_{1+(k-1)q}^m, a_{t_{k-1}+q}^m - a_{1+(k-1)q}^m)|,
\end{eqnarray}
where $w$ is a weighted parameter for the deviation angle $\theta$.
\item Compute and store the divergence rate of the $k$\textit{th} loop defined as $g_k(q)=d_2{(k)}/d_1{(k)}$, where $d_1(k) = \| a_{t_k}^m - a_{1+(k-1)q}^m \|$, $d_2(k) = \| a_{t_k+q}^m - a_{1+kq}^m \|$.
\end{enumerate}
\item Step 5: Finally, compute LLE $\lambda_q$ by averaging all $g_k(q)$ as $\lambda_q = \frac 1 K \Sigma_{k=1}^K [ln(d_2^{(k)}/d_1^{(k)})/q]$.
\end{itemize}
We stress that all LLE algorithms assume \textit{a priori} the values of the embedding dimension $m$ and the time delay $\tau$. These values are generally not known in advance, given a time series of a state output. While one can always search for $(m,\tau)$ using existing algorithms, it should be noted that the above LLE's algorithm will converge to a correct LLE of a chaotic attractor with a fractal dimension $n$, if it exists, for $m>2n+1$ as proven in Brock \cite{Brock1986}. As such, one can plot $\lambda(q)$ as a function of $(m,\tau)$ and search for the values of $(m,\tau)$ for which LLE becomes stabilized. This approach of searching for a LLE in the parameter space ($m,\tau$) is chosen in this study, because it can help reduce various prescribed thresholds in the current algorithms for $(m,\tau)$.         

Along with the LLE calculation, the method of detecting chaos proposed by Sugihara and May \cite{SugiharaMay1990} is also of particular interest and common because of its simplicity and effectiveness. The main idea behind this approach is that a chaotic time series should possess limited predictability, whereas true stochastic variation would have no predictability. Practically, this important property of chaotic time series implies that the correlation between model forecast and observations must decay with time in a chaotic system. 

For the sake of completeness, we summarize here the main step to obtain the Sugihara-May correlation as a function of forecast lead time ($T$) from a given time series. 
\begin{itemize}
\item Step 1: Given a time series $\{a_i\}, i=1...N$, one first divides it into an "atlas" (or training) set $\mathcal{A}$ and a test set $\mathcal{T}$; 
\item Step 2: Reconstruct a phase space with a given embedding dimension $m$ by generating the $m$-histories obtained from lagged time series as $a_t^m=(a_t,a_{t-\tau},\dots, a_{t-(m-1)\tau})$ for both sets $\mathcal{A},\mathcal{T}$; 
\item Step 3: For each history $a_i^m \in\mathcal{T}$ (the so-called predictee in Sugihara and May) in the $m$-dimensional space, search for $n_b$ neighbouring points in $\mathcal{A}$ with the minimal distance to $a_i^m$ such that the predictee is within a smallest simplex spanned by these $n_b$ neighbouring points; 
\item Step 4: a prediction at a given lead time $T$ for $a_i^m$ is then obtained by projecting the entire simplex into the future for $T$ leading time steps. The prediction value $a_i^f(T)$ at lead time $T$ for $a_i^m$ is then computed by taking an ensemble average of $n_b$ values of the simplex at lead time $T$; 
\item Step 5: construct a pair between the prediction $a_i^f(T)$ and the actual value of $a_i^m$ evolution after $T$ steps forward that is obtained directly from the training set $\mathcal{T}$, i.e., $a_{i+T}^m \in \mathcal{T}$;
\item Step 6: Repeat Steps 3-5 for all data points $a_i \in \mathcal{T}$ and obtain the correlation $\rho(T)$ between $(a^f_i(T),a^m_{i+T})$ for each lead time $T$;
\item Step 7: Repeat Steps 3-6 for different values of $T$ to obtain the curve $\rho(T)$ as a function of $T$. 
\end{itemize}
Note that in Step 4 of the above SMC algorithm, there are several different ways to obtain $a_i^f(T)$ (also known as "the prediction model") such as weighted average, regression combination, ensemble average, or neural network. Regardless of the prediction model, the key property of any chaotic time series is that $\rho(T)$ will decay with lead time $T$, which should remain valid in the presence of low-dimensional chaos. In this regard, the SMC curve $\rho(T)$ comprises a criterion for detecting chaotic time series; a deterioration of SMC with the leading time indicates the existence of chaos, whereas a purely stochastic time series would have a constant SMC regardless of how far into the future. More verification and applications of SMC for different systems can be found in \cite{SugiharaMay1990, Sugihara_etal1994}.     

Similar to the LLE algorithm, both the embedding dimension $m$ and the delay time $\tau$ have to be given before computing SMC. Our proposed approach to this freedom in choosing these parameters is to generate an SMC curve $\rho(T)$ for a range of values of $(m,\tau)$, as for the LLE analyses. The convergence of the SMC curve for some domain in the $(m,\tau)$ parameter space will then indicate the existence of a low-dimensional chaotic attractor in the embedding phase space. By comparing the values of $(m,\tau)$ obtained from the convergence of the SMC curves to the values of $m,\tau$ obtained from the convergence of LLE, one can estimate a proper range for $(m,\tau)$ that represents the chaotic regime of hurricane intensity, which are the main results presented in this study. More in-depth discussion about other methods for choosing optimal parameters $(m,\tau)$ can be found in Grassberger et al. \cite{Grassberger_etal1991}.  
%
%
\subsection*{Idealized hurricane simulations}
Given the approaches to detect chaos from time series described in the previous section, our next step is to generate a time series of hurricane intensity for the phase-space reconstruction analysis. In principle, one could obtain this time series directly from observation such as flight level data or satellite imagery. However, the requirement of a stationary time series for the phase-space reconstruction imposes a strong constraint on possible choices of time series, as observations of hurricane intensity contain various stages of hurricane development in different environments instead of just the mature stage. As such, using a hurricane model to produce the intensity time series in a fixed environment is the most apparent approach for our purpose. Ideally, one should use a full-physics three-dimensional model such as the Hurricane Weather Research and Forecasting model. These types of limited-area models are, nevertheless, designed on a rectangle domain with strong constraints by lateral boundary conditions, which prevent one from running for a very long time to generate a stationary time series. Because of this, we use an axisymmetrical model herein, which allows for a long integration without issue of lateral boundary asymmetries. 

In this study, the axisymmetric option of the cloud model (CM1,\cite{BryanFritsch2002}) was used to produce different intensity time series from a range of idealized hurricane simulations. The model was configured with 359 grid points on a stretching grid in the radial direction with the highest resolution of 2 km in the storm central region and stretched to 6 km outside 1000 km radius. In the vertical direction, a setting of 61 levels with a fixed resolution of 0.5 km was chosen. The model was initialized from the tropical Jordan sounding on an $f$-plane, with fixed sea surface temperature (SST) = 302.15 K. 

Because of the requirement of a quasi-stationary time series at the maximum intensity equilibrium, the model was configured for 100-day simulations. A stable configuration for this 100-day integration could be obtained by using a suite of physical paramterizations including the YSU boundary layer scheme, the TKE subgrid turbulence scheme, and  explicit moisture Kessler scheme with no cumulus parameterization. For the radiative parameterization, an idealized option with the Newtonian cooling relaxation of 2 K day$^{-1}$ was applied, similar to what used in \cite{KieuMoon2016}. This choice of the radiative cooling parameterization is sufficient to allow for a stable maximum intensity equilibrium during the entire 100-day simulations as shown in Figure \ref{fig1}. Given this stable configuration of hurricane intensity, the time series of $U_{MAX}$, $V_{MAX}$, $W_{MAX}$, and $P_{MIN}$ were then output at a sampling frequency of 36 seconds to optimize our time series analyses. 

As a step to verify the effects of random noise on our time series analyses, a set of sensitivity experiments were also conducted for which random white noise with a given variance was added to the CM1 model forcing at every time step. This implementation of additive random noise turns the CM1 model into a stochastic system whose output now contains random fluctuations with an amplitude proportional to the magnitude of random forcing. As discussed in \cite{Nguyen_etal2020}, this additive random noise in terms of the Wiener process with a constant variance results in a first-order accuracy for the CM1 model finite difference similar to the Euler-Maruyama method. By choosing a sufficiently small time step, the model is able to maintain its numerical stability for a range of experiments. Note that random noise was applied only to wind components at all model grid points, with an variance in the range of $[10^{-3}- 10^{-2} m s^{-1}]$. Beyond this range, we notice that the model violates the CFL conditions and quickly loses its stability after just a few steps of integration. Our rationale for applying random noise to the wind field in these sensitivity experiments is because wind components generally most fluctuate with time at any grid point. Adding random noises to the model temperature, pressure, and moisture fields does not change the outcomes, yet these extra noises would cause the model to become easily unstable and limit the range of random noise amplitude that we can implement for the wind components. Thus, all stochastic integration was carried out only for the wind components in this study.     



\section*{Data availability}
All data in this study can be generated by using the CM1 model settings described in the Method section of this study. 

\section*{Code availability}
The CM1 model is freely available at https://www2.mmm.ucar.edu/people/bryan/cm1/. All other phase-space reconstruction analyses are available from the corresponding author upon request.

\bibliography{reference_tc.bib}

\section*{Acknowledgements}
This research is partially supported by ONR/TCRI program (N000142012411).

\section*{Competing interests}
The authors hereby declare no conflict of interest in this study.

\section*{Author contributions statement}
CK perceived the idea, designed experiments, and wrote the manuscript. LF carried out mathematical analyses and contributed to editing the manuscript. WC performed simulations and analyses.

\section*{Figures \& Tables}
%
%
%
%
%
%
%
%
%
%


\end{document}